\documentclass[aps, twocolumn, showpacs]{revtex4}
\usepackage{graphicx}
\usepackage{dcolumn}
\usepackage{bm}
\usepackage{amsmath,amssymb}

\newcommand{\oper}[2]{\left | #1 \right\rangle\left\langle #2 \right|} 
\newcommand{\ket}[1]{\left | #1 \right\rangle}
\newcommand{\bra}[1]{\left\langle #1 \right|}
\newcommand{\avg}[1]{\left\langle #1 \right\rangle} 

\begin{document}

\title{Density Matrix Spectra and Order Parameters in the 1D Extended Hubbard Model}
\author{Wing Chi Yu}
\affiliation{Department of Physics and ITP, The Chinese University of Hong Kong, Hong
Kong, China}
\author{Shi-Jian Gu}
\altaffiliation{Email: sjgu@phy.cuhk.edu.hk}
\affiliation{Department of Physics and ITP, The Chinese University of Hong Kong, Hong
Kong, China}
\author{Hai-Qing Lin}
\affiliation{Beijing Computational Science Research Center, Beijing 100084, China}

\begin{abstract}
Without any knowledge of the symmetry existing in the system, we
derive the exact forms of the order parameters which show long-range
correlation in the ground state of the one-dimensional extended Hubbard model
using a quantum information approach. Our work demonstrates that the
quantum information approach can help us to find the explicit form
of the order parameter, which cannot be derived systematically via
traditional methods in the condensed matter theory.
\end{abstract}

\maketitle








\section{Introduction}

In many-body physics, the correlation function plays a fundamental
role. Especially in theoretical studies, the understanding on many
physical phenomena are based on the calculation of the corresponding
correlation functions\cite{Mahan,Sachdev}. For instance, to
investigate the magnetic properties of a system, people often
calculate the spin-spin correlation function to learn the possible
magnetic order. In 1D case, the system has a ferromagnetic order if
the $0$ mode is dominant, and an anti-ferromagnetic order if the
$\pi$ mode is dominant. The long-range behavior of the correlation
function is associated with the symmetry-breaking in the system,
which is an important concept in the understanding of continuous
phase transitions. That is, for a certain operator, its
non-vanishing value at a long distance denotes a symmetry-broken
phase which usually occurs in the thermodynamic limit.
Mathematically, if the correlation function
\begin{eqnarray}
C(0,r)=\left\langle O_{0}O_{r}\right\rangle -\left\langle O_{0}\right\rangle
\left\langle O_{r}\right\rangle
\label{eq:corr}
\end{eqnarray}
is a constant in the infinite $r$ limit, we say that the state has a
long-range order. The operator $O$ can then be taken as the order
parameter to describe the corresponding phase.

Traditionally, to find an appropriate order parameter for a certain
phase, people often resort to other methods such as group theory and
the renormalization group analysis. Nevertheless, these methods
cannot always help us to find the correct form of the order
parameter. Hence a method to derive the order parameter
systematically is very instructive. Using the variational approach,
Furukawa \emph{et al}\cite{DriveOrder} proposed a scheme to derive
the order parameter. While the scheme is promising, it still needs
the knowledge of the degenerated states that lead to the symmetry
breaking in the thermodynamic limit. Instead, we proposed recently a
scheme to derive the order parameter from the spectrum of the
reduced density matrix of the ground state directly \cite{COrder}.
Differ from Furukawa's scheme, our approach is not
variational, and needs only the knowledge of the ground state.

In this paper, we apply the scheme, for the first time, to a realistic
model. To start with, we will first demonstrate the derivation of the order parameter in the Hubbard model in section \ref{sec:HM}. Then, we apply our scheme to the extended Hubbard model (EHM) in section \ref{sec:EHM} and \ref{sec:EHM2}. We show that even without any knowledge of the symmetry of the system, we can derive
the exact forms of the order parameters which show long-range
correlations in the ground state of the model. A summary would be given in section \ref{sec:sum}.

\begin{figure}[tbp]
\includegraphics[width=8.1cm]{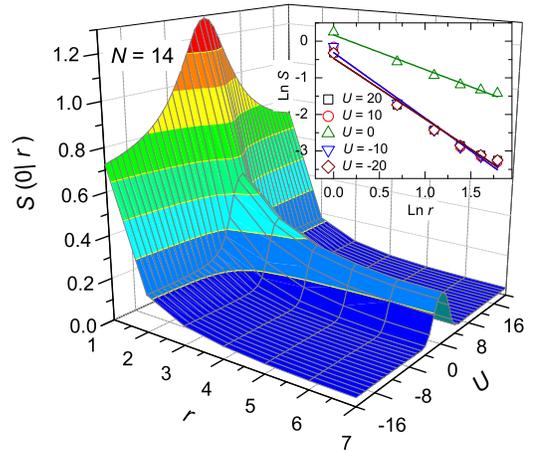}
\caption{(color online) The mutual information as a function of $r=|i-j|$
and $U$ in the Hubbard model. The inset shows that the mutual information decays algebraically with $r$ and thus long-range correlation is presented in the system.} \label{figure_mutualinformation}
\end{figure}

\section{Deriving the order parameters}
The Hubbard type models has been served as a prototype in condensed
matter physics to study the electron correlation effect in solids.
Especially in many cases where the long-range Coulomb interaction
has significant importance, like in quasi one-dimensional organic
solids such as conjugated polymers \cite{AP1} and charge transfer
crystals \cite{AP2}, the extended Hubbard model in one-dimension is
the simplest model that included finite range interactions for
theoretical studies.
 The model's Hamiltonian reads%
\begin{eqnarray}
H&=&-t\sum_{\sigma ,j} \left(c_{j,\sigma }^{\dagger
}c_{j+1,\sigma}+h.c.\right)\nonumber\\
&&+U\sum_{j}n_{j,\uparrow }n_{j,\downarrow }+V\sum_{j}n_jn_{j+1},
\end{eqnarray}%
where $c_{j,\sigma }^{\dagger }$ and $c_{j,\sigma }$ ($\sigma
=\uparrow ,\downarrow $) are creation and annihilation operators of
electrons with a spin $\sigma $ at the site $j$ respectively, $n_{j,\sigma}=c_{j,\sigma}^{\dagger }c_{j,\sigma }$, and $n_j=n_{j,\uparrow}+n_{j,\downarrow}$. $U$ and $V$ is the
strength of the on-site and the nearest-neighbor Coulomb interaction respectively. $t$ is the hoping integral
and is taken to be unity for convenience.

Interestingly, the EHM exhibits a very rich phase diagram. It was
shown analytically \cite{T1,T2,T3,T4} and numerically \cite{N1, N2}
the existence of spin-density waves (SDW), charge-density waves
(CDW), phase separation (PS), singlet superconducting (SS) and the
triplet superconducting (TS) phases in the model. Using concepts
from quantum information theory as a tool, it is also shown that the
local entanglement \cite{EHM} was able to obtain the CDW, SDW and PS
phases. In addition to these three phases, the block-block
entanglement can also witness the SS phase \cite{Deng}. Besides the
phases mentioned above, there are also studies pointed out the existence of the bond-order waves
phase(BOW) in the model. However, whether this BOW exists as a narrow region or just a line in the ground state phase diagram is still a controversial problem \cite{BOW, Jeckelmann, Jeckelmann2} .

At this point, let us first forget about the result from previous studies and suppose we know nothing about the nature of the phases.
To find the potential
order existing in the ground state of the model, we need to first
examine if the ground state exist a long-range correlation or not. While we
do not know the form of any order parameters, we can still use the
mutual information, a concept borrowed from quantum information
science, to measure the total correlation between two arbitrary
blocks in the system. The mutual information is defined as%
\begin{equation}
S(i|j)=S\left( {\rho }_{i}\right) +S\left( {\rho }_{j}\right) -S\left( {\rho
}_{i\cup j}\right) ,
\end{equation}%
where $S\left( {\rho }_{i}\right) =-$tr$({\rho }_{i}\log _{2}{\rho
}_{i})$ is the von-Neumann entropy of the block $i$. $\rho_i$ is the reduced density matrix obtain by tracing out all other degrees of freedom except that of site $i$, i.e.${\rho }_{i}=$tr$\left\vert \Psi _{0}\right\rangle \left\langle \Psi_{0}\right\vert $ where $\ket{\Psi_0}$ is the ground state of the system. It has been shown that the existence of the long-range
order means that there is a non-vanishing mutual information at a
long distance \cite{MMWolf,SJGuJPA}. This statement still holds true
vice versa. By studying the spectra of the reduced density
matrices, we can derive the potential order parameters.

\subsection{Case I: $V=0$}
\label{sec:HM} To illustrate the scheme more explicitly, let us
first ignore the next-nearest neighbor interaction, i.e. $V=0$, and
the model is reduced to the conventional Hubbard model
\cite{Hubbdard, Gutzwiller, Kanamori}.

Consider a block consists of one site, i.e. $N_B=1$. In the basis of local states $\{\ket{\mu}\}=\{\left\vert
0\right\rangle ,\left\vert \uparrow \right\rangle ,\left\vert
\downarrow \right\rangle ,\left\vert \uparrow \downarrow
\right\rangle \}$, ${\rho }_{i}$ is
found to be diagonal, i.e.
\begin{equation}
\rho _{i}=u\left\vert 0\right\rangle \left\langle 0\right\vert +v\left\vert
\downarrow \right\rangle \left\langle \downarrow \right\vert +v\left\vert
\uparrow \right\rangle \left\langle \uparrow \right\vert +u\left\vert
\uparrow \downarrow \right\rangle \left\langle \uparrow \downarrow
\right\vert,
\label{eq:rhoi}
\end{equation}
where $u$ and $v$ are some positive real numbers.
The two-site reduced density matrix ${\rho }_{i\cup j}$ is a
block-diagonal matrix in the basis $\{\ket{\mu\nu}\}=\{\left\vert 0\right\rangle
,\left\vert \uparrow \right\rangle ,\left\vert \downarrow
\right\rangle ,\left\vert \uparrow \downarrow \right\rangle
\}\otimes \{\left\vert 0\right\rangle ,\left\vert \uparrow
\right\rangle ,\left\vert \downarrow \right\rangle ,\left\vert
\uparrow \downarrow \right\rangle \}$. Figure 1 shows the dependence
of the mutual information $S(0|r)$ as a function of $r=|i-j|$ and
$U$ calculated for a 14-site system with 7 spin-up electrons and 7
spin-down electrons. The results are obtained from numerical exact diagonalization
under periodic boundary conditions. From the
figure, we can see that the mutual information reaches a maximum at
$U=0$, which is a non-trivial point of the Hubbard model. We will
return to this phenomenon later. At this stage we are more
interested in the long-range behavior of the mutual information. For
this purpose, we plot the dependence of $S(0|r)$ on $r$ in the inset
of Figure 1 for various $U$. From the inset, it is easy to find that $%
S(0|r) $ decays algebraically with $r$. Therefore, we judge that in
the ground state of the 1D Hubbard model, there must exist certain
kind of long-range correlation though we do not know the explicit
form of the order parameter yet.

\begin{figure}[tbp]
\includegraphics[width=9cm]{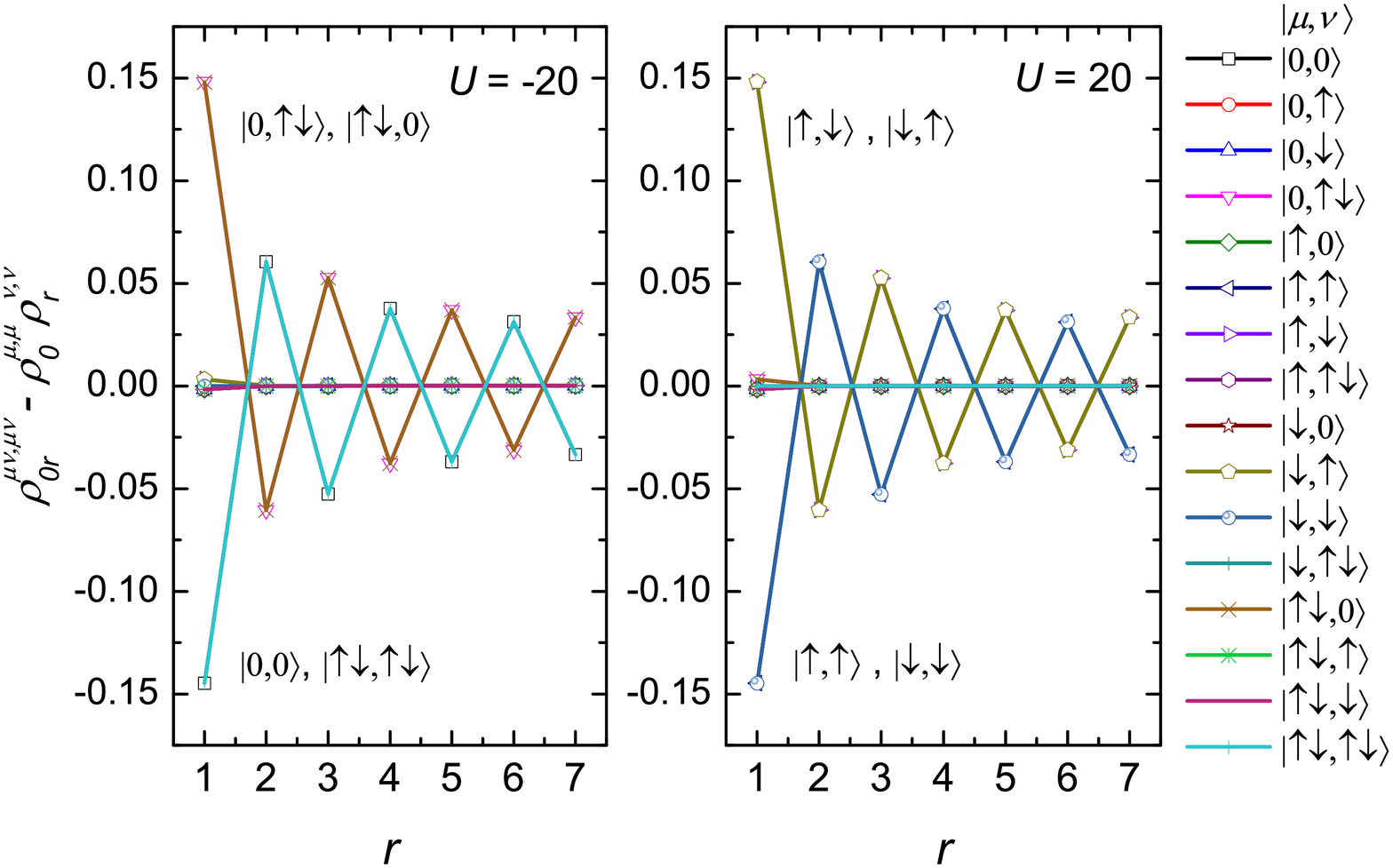}
\caption{(color online) A plot of
$\rho_{0r}^{\mu\nu,\mu\nu}-\rho_0^{\mu,\mu}\rho_r^{\nu,\nu}$ as a
function of $r$ for two limiting cases of $U=-20, 20$
in the Hubbard model. For $U=-20$ (left), the charge degrees of
freedom ($\{\mu,\nu\}\in\{\left\vert
0\right\rangle, \left\vert \uparrow \downarrow
\right\rangle\}$) are correlated, while for $U=20$(right) the
spin degrees of freedom($\{\mu,\nu\}\in\{\left\vert \uparrow \right\rangle ,\left\vert
\downarrow \right\rangle \}$) are correlated. } \label{figure_ev_rdm1dddd}
\end{figure}

\begin{figure}[tbp]
\includegraphics[width=8.1cm]{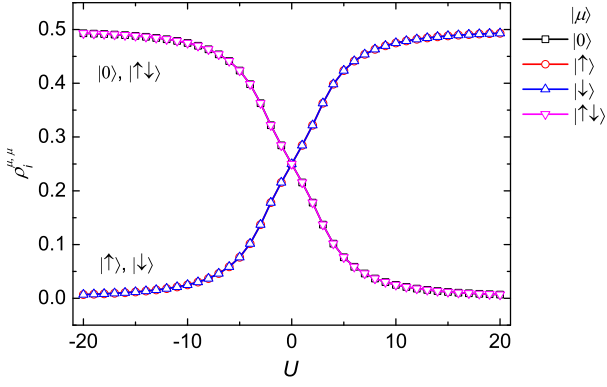}
\caption{(color online) The diagonal matrix elements, which is also the
eigenvalues, of $\protect\rho_i$ as a function of $U$ in the Hubbard
model. Here $N=14$ and $N_B=1$. For the negative $U$ case, the
eigenstates $|0\rangle$ and $|\uparrow\downarrow\rangle$ have
dominate weights while for positive $U$, the eigenstates
$|\uparrow\rangle$ and $|\downarrow\rangle$ have dominate weights.} \label{figure_ev_rdm1}
\end{figure}

\textit{Diagonal order parameter:} According to the scheme we
proposed recently\cite{COrder}, the order parameter can be
constructed from the spectrum of the reduced density matrix $\rho
_{i}$. However, in most of the cases, we cannot obtain results of
the total correlation between two blocks separated by a large enough
distance due to the limitation of computational powers.

To single out the correlated elements in $\rho _{i}$, we calculated the difference between the
diagonal matrix elements of $\rho_{i\cup j}$ and that of
$\rho_i\otimes\rho_j$ as a function of the separation $|i-j|$.
Figure \ref{figure_ev_rdm1dddd} shows a plot of $\rho _{0r}^{\mu\nu,\mu\nu}-\rho _{0}^{\mu,\mu}\rho
_{r}^{\nu,\nu}$ as a function of $r$. For convenience, the two block reduced density
matrix $\rho_{0\cup r}$ is written as $\rho_{0r}$. From the figure,
clearly we can see that $\rho_{0r}\ne\rho_0\otimes\rho_r$ which is
consistence with $S(0|r)\ne 0$. Moreover, for $U=-20$, the
difference in the diagonal matrix elements for
$\{\mu,\nu\}\in\{\left\vert 0\right\rangle, \left\vert \uparrow
\downarrow \right\rangle\}$ decays algebraically with $r$ while it
is almost zero for the others. We can then argue that the vacant
state and the double occupancy state of the two blocks are
correlated in this case. For the case of $U=20$, $\rho
_{0r}^{\mu\nu,\mu\nu}-\rho _{0}^{\mu,\mu}\rho _{r}^{\nu,\nu}$ are
almost zero unless $\{\mu,\nu\}\in\{\left\vert \uparrow
\right\rangle ,\left\vert \downarrow \right\rangle \}$. In this
limit, the spin up and spin down states are correlated.

\begin{figure}[tbp]
\includegraphics[width=8.5cm]{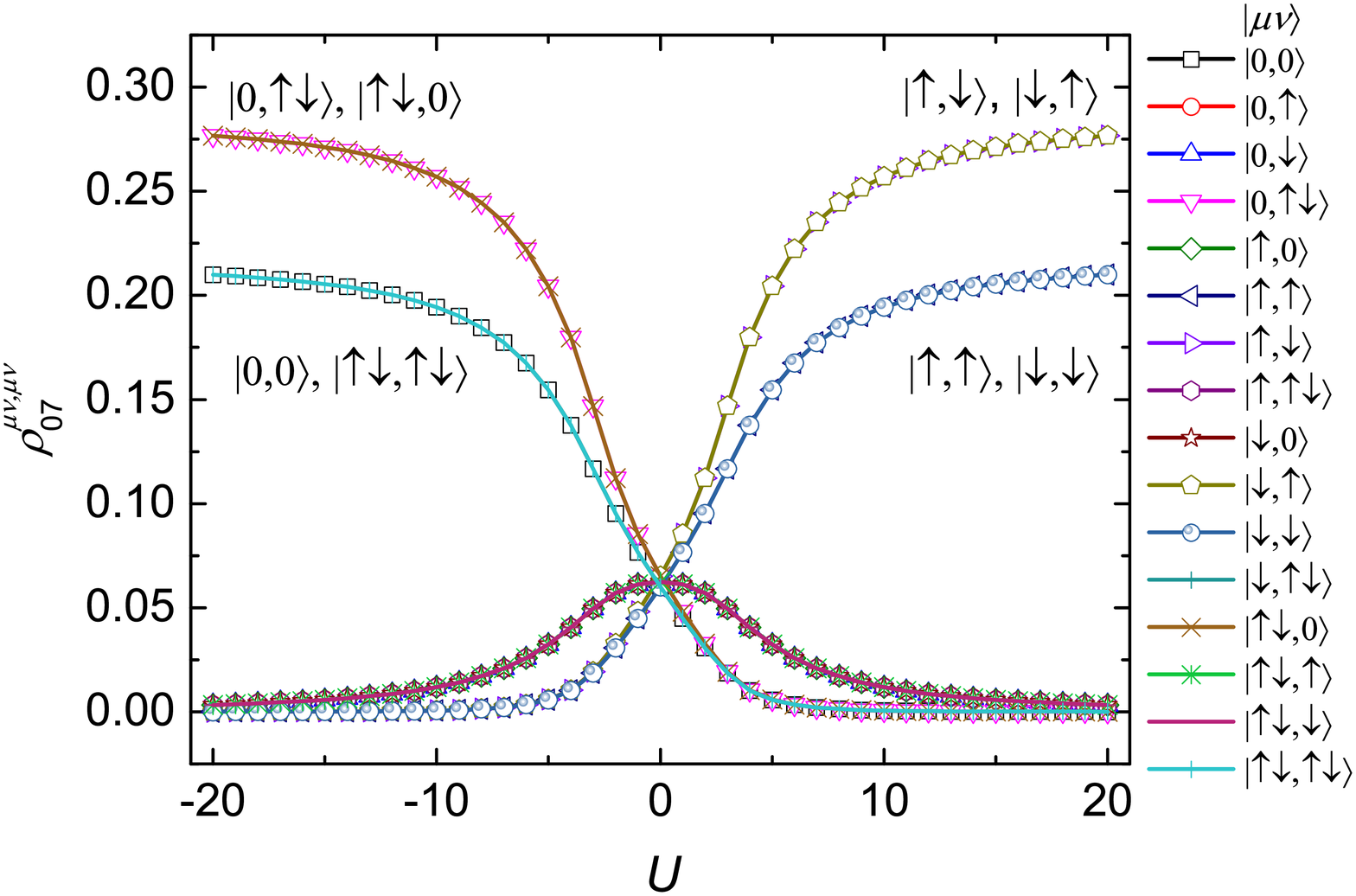}
\caption{(color online) The diagonal elements of $\protect\rho_{ij}$
as a function of $U$ in the Hubbard model for $r=|i-j|=7$. In the negative
$U$ limit, only the matrix elements corresponds to the basis $\{\mu,
\nu\}\in\{|0\rangle,|\uparrow\downarrow\rangle\}$ is non-vanishing.
In the positive $U$ limit, only the matrix elements corresponds to
the basis $\{\mu, \nu\}\in\{|\uparrow\rangle,|\downarrow\rangle\}$
is non-vanishing.} \label{figure_rdm2_r7_dia}
\end{figure}

\begin{figure}[tbp]
\includegraphics[width=8.5cm]{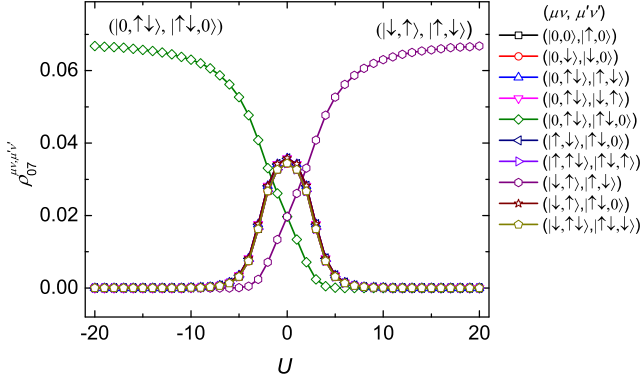}
\caption{(color online) The non-zero off-diagonal elements of $\protect\rho%
_{ij}$ as a function of $U$ in the Hubbard model for $r=7$. The
matrix elements for $(\mu\nu,\mu'\nu')$ is the same as
$(\mu'\nu',\mu,\nu)$.} \label{figure_rdm2_r7_offdia}
\end{figure}

In Figure \ref{figure_ev_rdm1}, we show the dependence of the eigenvalues of $%
\rho _{i}$ on the interaction $U$. From the figure, we can see that
if $U$ becomes larger and larger, the eigenvalues of the state
$\left\vert \downarrow \right\rangle$ and $\left\vert \uparrow
\right\rangle $ tend to 1/2 and the eigenvalues of $\left\vert
0\right\rangle$ and $\left\vert \uparrow \downarrow \right\rangle $
vanish. The observation means that in the large $U$ limit, the
reduced density matrix becomes
\begin{equation}
\rho _{i}=\frac{1}{2}\left\vert \downarrow \right\rangle \left\langle
\downarrow \right\vert +\frac{1}{2}\left\vert \uparrow \right\rangle
\left\langle \uparrow \right\vert .
\end{equation}%
In the basis of $\rho _{i}\otimes\rho _{j}$, we show the non-zero
diagonal and off-diagonal elements of ${\rho }_{i\cup j}$ in Figure
\ref{figure_rdm2_r7_dia} and Figure \ref{figure_rdm2_r7_offdia}
respectively. From the figures, we find that $\rho _{0r}$ takes the
form
\begin{equation}
\rho _{0r}=\left(
\begin{array}{cccc}
u & 0 & 0 & 0 \\
0 & v & z & 0 \\
0 & z & v & 0 \\
0 & 0 & 0 & u%
\end{array}%
\right)   \label{eq:singlerdm}
\end{equation}%
in the large $U$ limit in the basis of $\{\left\vert \uparrow \right\rangle
\left\vert \uparrow \right\rangle ,\left\vert \downarrow \right\rangle
\left\vert \uparrow \right\rangle ,\left\vert \uparrow \right\rangle
\left\vert \downarrow \right\rangle ,\left\vert \downarrow \right\rangle
\left\vert \downarrow \right\rangle \}$. Therefore,
the order parameter can be written as
\begin{equation}
O^{d}_i=w_{1}\left\vert \uparrow \right\rangle \left\langle \uparrow
\right\vert +w_{2}\left\vert \downarrow \right\rangle \left\langle
\downarrow \right\vert .
\end{equation}%
According to the traceless condition tr$(O_{i}^{d}\rho _{i})=0$, we have $%
w_{1}=-w_{2}$. Let $w_{1}=1$, then the order parameter becomes%
\begin{equation}
O_{i}^{d}=\left\vert \uparrow \right\rangle \left\langle \uparrow
\right\vert -\left\vert \downarrow \right\rangle \left\langle \downarrow
\right\vert .
\end{equation}%
For simplicity, we denote it as $\sigma _{i}^{z}$.

Similarly, for the case of negative $U$, we can find that
\begin{equation}
\rho _{i}=\frac{1}{2}\left\vert 0\right\rangle \left\langle 0\right\vert +%
\frac{1}{2}\left\vert \downarrow \uparrow \right\rangle \left\langle
\downarrow \uparrow \right\vert .
\end{equation}%
and
\begin{equation}
\rho _{0r}=\left(
\begin{array}{cccc}
u & 0 & 0 & 0 \\
0 & v & z & 0 \\
0 & z & v & 0 \\
0 & 0 & 0 & u%
\end{array}%
\right)
\label{eq: rho_charge}
\end{equation}%
in the basis of $\{\left\vert 0\right\rangle \left\vert 0\right\rangle
,\left\vert 0\right\rangle \left\vert \downarrow \uparrow \right\rangle
,\left\vert \downarrow \uparrow \right\rangle \left\vert 0\right\rangle
,\left\vert \downarrow \uparrow \right\rangle \left\vert \downarrow \uparrow
\right\rangle \}$. Then the diagonal order parameter can be defined as%
\begin{equation}
O_{i}^{d}=w_{1}\left\vert 0\right\rangle \left\langle 0\right\vert
+w_{2}\left\vert \downarrow \uparrow \right\rangle \left\langle \downarrow
\uparrow \right\vert .
\end{equation}%
Apply the traceless condition and cut-off condition, we get
\begin{equation}
O_{i}^{d}=\left\vert 0\right\rangle \left\langle 0\right\vert -\left\vert
\downarrow \uparrow \right\rangle \left\langle \downarrow \uparrow
\right\vert .
\end{equation}%
For simplicity, we denote it as $\eta _{i}^{z}$.

\textit{Off-diagonal order parameter:} From Eq.
(\ref{eq:singlerdm}), we find that $\rho _{0r}$ is not a diagonal
matrix. This observation means that there exists off-diagonal
long-range correlations in the ground state of the 1D Hubbard model.
According to our scheme, the off-diagonal order parameter can be
defined as
\[
O_{i}^{o}=w\left\vert \uparrow \right\rangle \left\langle \downarrow
\right\vert +w^{\ast }\left\vert \downarrow \right\rangle \left\langle
\uparrow \right\vert .
\]%
Here $w$ is complex, so there are two independent parameters in the
operator. We separate the real part and the imaginary part in the
operator
\begin{eqnarray}
O_{i}^{o} &=&x\left( \left\vert \uparrow \right\rangle \left\langle
\downarrow \right\vert +\left\vert \downarrow \right\rangle \left\langle
\uparrow \right\vert \right) +iy\left( \left\vert \uparrow \right\rangle
\left\langle \downarrow \right\vert -\left\vert \downarrow \right\rangle
\left\langle \uparrow \right\vert \right)  \nonumber \\
&=&xO_{i}^{x}-yO_{i}^{y}.
\end{eqnarray}%
with
\begin{eqnarray*}
O_{i}^{x} &=&\left\vert \uparrow \right\rangle \left\langle \downarrow
\right\vert +\left\vert \downarrow \right\rangle \left\langle \uparrow
\right\vert , \\
O_{i}^{y} &=&-i(\left\vert \uparrow \right\rangle \left\langle
\downarrow \right\vert -\left\vert \downarrow \right\rangle
\left\langle \uparrow \right\vert ).
\end{eqnarray*}%
Obviously, we have
\begin{eqnarray}
\avg{O_i^x}=\avg{O_i^y}=0,
\end{eqnarray}
and
\begin{eqnarray}
\avg{O_i^xO_j^x}=\avg{O_i^yO_j^y}=2z.
\end{eqnarray}
So we can treat either $O_i^x$ or $O_i^y$ or their linear combination as the off-diagonal order operator.
Let's denote them as $\sigma _{i}^{x}$ and $\sigma _{i}^{y}$ respectively.

Similarly, in the negative $U$ region, we can also derive the
off-diagonal order
parameter as%
\begin{eqnarray*}
O_{i}^{x} &=&\left\vert 0\right\rangle \left\langle \uparrow \downarrow
\right\vert +\left\vert \uparrow \downarrow \right\rangle \left\langle
0\right\vert , \\
O_{i}^{y} &=&-i(\left\vert 0\right\rangle \left\langle \uparrow
\downarrow \right\vert -\left\vert \uparrow \downarrow \right\rangle
\left\langle 0\right\vert ).
\end{eqnarray*}%
We denote them as $\eta _{i}^{x}$ and $\eta _{i}^{y}$ respectively.

$SU(2)$ \textit{algebra and the effective Hamiltonian: }In the basis of $%
\{\left\vert \uparrow \right\rangle ,\left\vert \downarrow \right\rangle \}$%
, $\sigma _{i}^{x},\sigma _{i}^{y}$ and $\sigma _{i}^{z}$ can be written as%
\begin{equation}
\sigma _{i}^{x}=\left(
\begin{array}{cc}
0 & 1 \\
1 & 0%
\end{array}%
\right) ,\sigma _{i}^{y}=\left(
\begin{array}{cc}
0 & -i \\
i & 0%
\end{array}%
\right) ;\sigma _{i}^{z}=\left(
\begin{array}{cc}
1 & 0 \\
0 & -1%
\end{array}%
\right) .
\end{equation}%
These expressions are exactly the form the Pauli matraces, which satisfy the
su(2) Lie algebra, i.e. $[\sigma _{i}^{\alpha },\sigma _{j}^{\beta
}]=2i\delta _{ij}\epsilon ^{\alpha \beta \gamma }\sigma _{i}^{\gamma }$. As
we know, in the large $U$ limit, the Hubbard model becomes the
one-dimensional Heisenberg model
\[
H=J\sum_{j}\mathbf{S}_{j}\cdot \mathbf{S}_{j+1},
\]%
where $J=4t^{2}/U$. In the ground state of the Heisenberg model, the
spin-spin correlation has a power-law decaying behavior. The order parameters derived is consistent with our knowledge.

On the other hand, in the negative $U$ region, $\eta _{i}^{x}$, $\eta
_{i}^{y}$, and $\eta _{i}^{z}$ can be written as%
\begin{equation}
\eta _{i}^{x}=\left(
\begin{array}{cc}
0 & 1 \\
1 & 0%
\end{array}%
\right) ,\eta _{i}^{y}=\left(
\begin{array}{cc}
0 & -i \\
i & 0%
\end{array}%
\right) ;\eta _{i}^{z}=\left(
\begin{array}{cc}
1 & 0 \\
0 & -1%
\end{array}%
\right)
\end{equation}%
in the basis of $\{\left\vert 0\right\rangle ,\left\vert \uparrow \downarrow
\right\rangle \}$. Clearly the three operators satisfy the su(2) Lie algebra
too.

\begin{figure}[tbp]
\includegraphics[width=8.1cm]{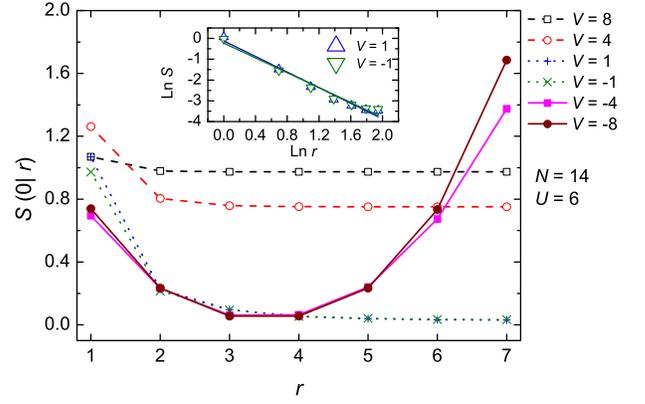}
\caption{(color online) The mutual information as a function of $r=|i-j|$ for $U=6$ in EHM. The inset shows a ln-ln plot of the mutual information versus $r$ for $V=\pm1$.} \label{figure_Entro_EHM}
\end{figure}

\begin{figure}[tbp]
\includegraphics[width=8.1cm]{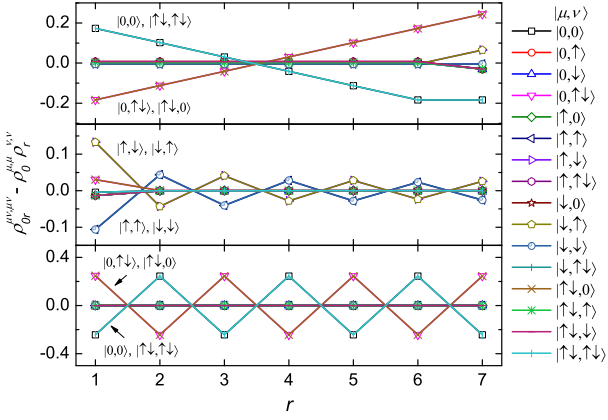}
\caption{(color online) A plot of
$\rho_{0r}^{\mu\nu,\mu\nu}-\rho_0^{\mu,\mu}\rho_r^{\nu,\nu}$ as a
function of $r$ for $V=-8$ (top panel), $V=1$ (middle panel), and
$V=8$ (bottom panel) in EHM for $U=6$. For $V=\pm 8$,
the charge degrees of freedom ($\{\mu,\nu\}\in\{\left\vert
0\right\rangle, \left\vert \uparrow \downarrow \right\rangle\}$) are
correlated, while for $V=1$ the spin degrees of
freedom($\{\mu,\nu\}\in\{\left\vert \uparrow \right\rangle
,\left\vert \downarrow \right\rangle \}$) are correlated.}
\label{figure_diff_EHM}
\end{figure}

\subsection{Case II: $V\neq 0$, $U=6$}
\label{sec:EHM}
Now, let us also include the next-nearest neighbor interaction in the system. Figure \ref{figure_Entro_EHM} shows a plot of the mutual information as a function of $r$. Obviously, the mutual information is non-vanishing at a large $r$ for $V=\pm 4, \pm8$ (in fact, they corresponds to two different regime where $V>U/2$ and $V<-U/2$ which we will see more explicitly later). For the intermediate case, we can see from the log-log plot in the inset that the mutual information shows a power law decaying behavior. One can then safely argue that the system exhibits certain kind of long-range correlations and we can go on to investigate the spectrum of the reduce density matrix to derive the potential order parameters.

Using the same basis as that for the Hubbard model in the previous subsection, we calculated the single-site and two-sites reduced density matrices. $\rho_i$ is diagonal and takes the form of Eq. \ref{eq:rhoi}. Figure \ref{figure_diff_EHM} shows a plot of the difference between the diagonal matrix element of $\rho_{0r}$ and the product of the diagonal matrix elements of $\rho_0$ and $\rho_r$ with the corresponding basis as a function of $r$. The finite difference in $\rho_{0r}^{\mu\nu,\mu\nu}-\rho_0^{\mu,\mu}\rho_r^{\nu,\nu}$ for some particular $\mu$ and $\nu$ indicates that $\rho_{0r}\ne\rho_0\otimes\rho_{r}$. For $V=-8$ and $V=8$, we see that the main contributors to the correlation of two sites separated at a large distance are the states $|0\rangle$ and $|\uparrow\downarrow\rangle$. While for the case of $V=1$, the $|\uparrow\rangle$ and $|\downarrow\rangle$ states play the role.

\begin{figure}[tbp]
\includegraphics[width=8.5cm]{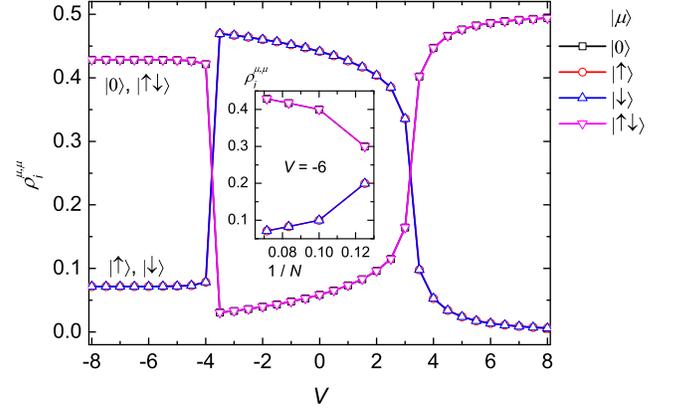}
\caption{(color online) A plot of the eigenvalues of $\rho_i$ as a
function of $V$ for $U=6$ in EHM. The inset shows the size
dependence of the eigenvalues of $\rho_i$ for $V=-6$. For $V\lesssim-U/2$ and $V>U/2$, the
eigenstates $|0\rangle$ and $|\uparrow\downarrow\rangle$ have
dominate weights. For $-U/2\lesssim V<U/2$, the eigenstates
$|\uparrow\rangle$, and $|\downarrow\rangle$ are dominated.}
\label{figure_EV1_EHM}
\end{figure}

Figure \ref{figure_EV1_EHM} plots the eigenvalues of the single-site reduced density matrix as a function of $V$. The crossings of the eigenvalues, which is the probabilistic weight of the corresponding eigenstates, of $\rho_i$ separate the system into three different regimes corresponding to $V\lesssim-U/2$, $-U/2\lesssim V< U/2$, and $V>U/2$. This echoes what we have mentioned in the previous paragraph. In each of the regime, the kind of correlation existing in the system is qualitatively different as one can judge from the nature of the dominating eigenstates.

For $-U/2\lesssim V< U/2$, the eigenvalues of the $|\downarrow\rangle$ and $|\uparrow\rangle$ states are around $1/2$ while that are almost zero for the $|0\rangle$ and $|\uparrow\downarrow\rangle$ states. Following similar argument in the previous section, we can define the order parameter as
\begin{eqnarray*}
O_i=w_1|\uparrow\rangle\langle\uparrow|+w_2|\downarrow\rangle\langle\downarrow|.
\end{eqnarray*}
Again, by applying the traceless condition and take $w_1=1$, we obtain the order parameter
\begin{eqnarray}
O_i=|\uparrow\rangle\langle\uparrow|-|\downarrow\rangle\langle\downarrow|=\sigma_i^z.
\end{eqnarray}

For $V\lesssim-U/2$ and $V>U/2$, the eigenvalues corresponds to the
states $|0\rangle$ and $\uparrow\downarrow\rangle$ are around
one-half while that for the states $|\uparrow\rangle$ and
$|\downarrow\rangle$ tends to zero. Note that the discrepancy from
the value $1/2$ and $0$ respectively for $V\lesssim-U/2$ is due to
the finite size effect. Nevertheless, we show the size dependence of
$\rho_i^{\mu,\mu}$ for $V=-6$ in the inset of Fig.
\ref{figure_EV1_EHM}. The eigenvalues for the $|\uparrow\rangle$ and
$|\downarrow\rangle$ becomes vanishing as the system size increases.
We can define the order parameter as
\begin{eqnarray*}
O_i=w_1|0\rangle\langle 0|+w_2|\uparrow\downarrow\rangle\langle\uparrow\downarrow|.
\end{eqnarray*}
The choice of the order parameter can be narrow down by applying the traceless condition and again taking $w_1$ to be $1$. We have
\begin{eqnarray}
O_i=|0\rangle\langle 0|-|\uparrow\downarrow\rangle\langle\uparrow\downarrow|=\eta_i^z.
\label{eq:O_CDW}
\end{eqnarray}

\begin{figure}[tbp]
\includegraphics[width=8cm]{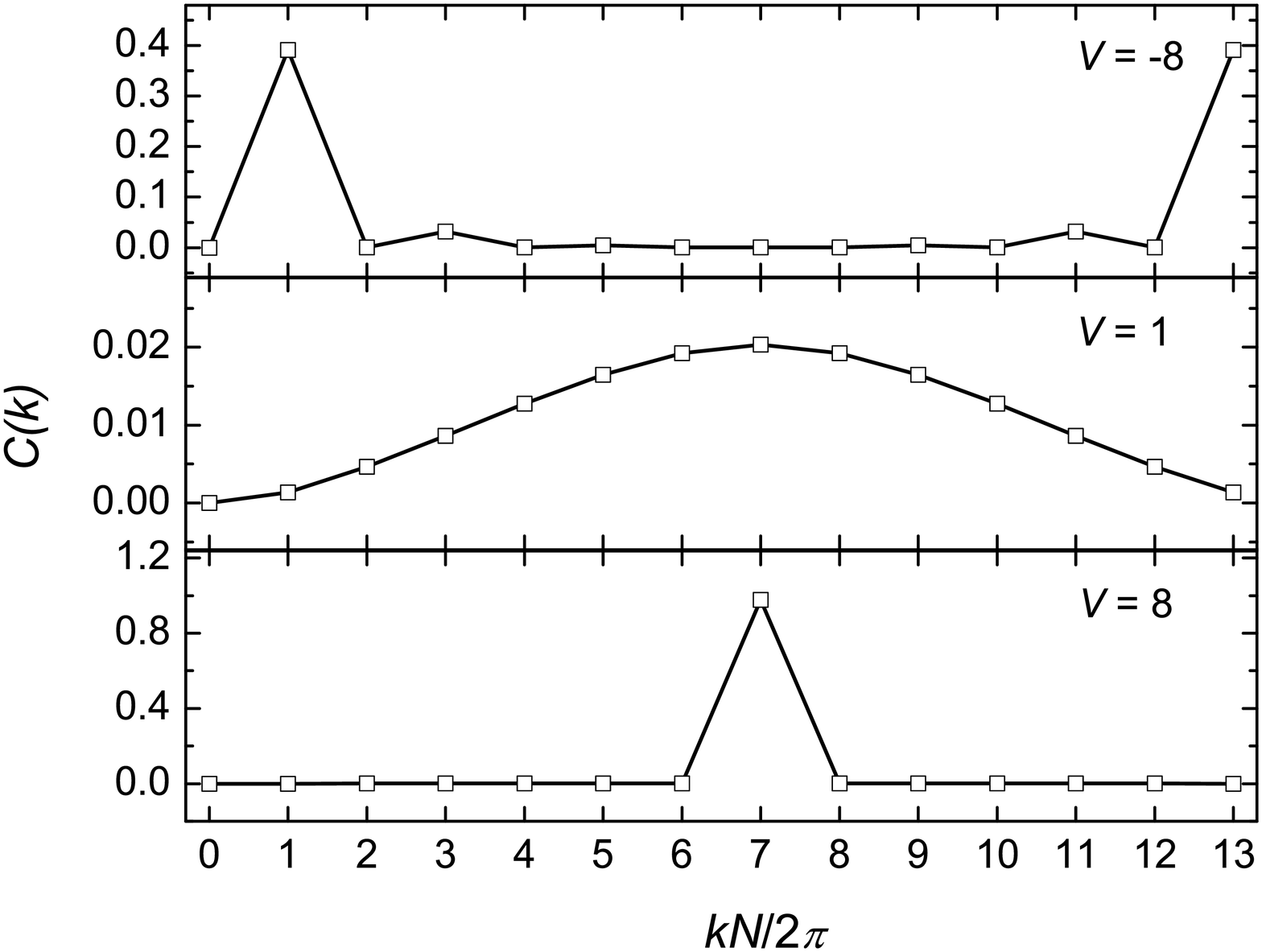}
\caption{The correlation function calculated from the derived order operators in the momentum space as a function of $k$ in EHM. Here $U=6$, $N=14$, and $V=-8$ (top panel), $V=1$ (middle panel), and $V=8$ (bottom panel) respectively.} \label{figure_cor_EHM}
\end{figure}

Here we obtained the same order parameter for both of the regimes where $V\lesssim-U/2$ and $V>U/2$. However, recall from Fig. \ref{figure_Entro_EHM}, the mutual information as a function of $r$ shows qualitatively different behavior in these two range of values of $V$. For $V>U/2$, the mutual information almost remains constant as $r$ increases. On the other hand, for $V\lesssim-U/2$, the mutual information first decreases and then increases to a maximum value upon reaching $r=7$, where it is the middle of the chain. It is reasonable to suspect that the ground state of the system is qualitatively different in these two regimes. To further distinguish between them, it is worth to study the mode of the order parameters.

With the order parameter found in Eq. (\ref{eq:O_CDW}), one can calculate the correlation function in Eq. (\ref{eq:corr}) and it is found to be oscillating with respect to $r$ as expected. To capture the mode of the order parameter, let's consider the Fourier transformation of the correlation function, i.e.
\begin{eqnarray}
C(k)=\frac{1}{N}\sum_{r=0}^{N-1}e^{-ikr} C(0,r)=\frac{1}{N}\sum_{r=0}^{N-1}e^{-ikr} \langle\eta_0^z\eta_r^z\rangle,
\end{eqnarray}
where $k=2m\pi/N$ and $m=0,1,\cdots,N-1$. Also note that $\langle\eta_i^z\rangle=0$ as a result of applying the traceless condition. The result is plotted in Fig. \ref{figure_cor_EHM}.

From the figure, we see that the correlation function peaks at $k=\pi$ for $V=8$. One can expect this holds true for the whole range of  $V>U/2$. Together with the form of the derived order parameter, we may conclude that the dominating configuration in the ground state of the system are consisting of alternating vacant and double occupancy states. This is in fact the well-know charge-density wave (CDW) states $|0,\uparrow\downarrow,\cdots,0,\uparrow\downarrow\rangle$ and $|\uparrow\downarrow,0,\cdots,\uparrow\downarrow,0\rangle$ in the extended Hubbard model.

For $V\lesssim-U/2$, the mode of the order parameter are $k=2\pi/N$
and $2\pi(N-1)/N$. The second peak in the correlation function
$C(k)$ is just a result from the periodic boundary conditions. In
this regime, one can deduce that the dominating ground state
configurations has a period of half of the lattice in the real
space. They are the phase separation (PS) states
$|0,\cdots,0,\uparrow\downarrow,\cdots,\uparrow\downarrow\rangle$
and the translation of it. This also explains why the mutual
information is maximum for the sites separated by half of the
lattice. Since any of the translation of the above state are equally
weighted in the ground state before symmetry is broken, only the
local states separated by half of the lattice can be confidently
determine once we know one of them. They have to be opposite to each
other. 

For completeness, we also studied the mode of the order parameter for $-U/2\lesssim V<U/2$. The correlation function
\begin{eqnarray}
C(k)=\frac{1}{N}\sum_{r=0}^{N-1}e^{-ikr} \langle\sigma_0^z\sigma_r^z\rangle
\end{eqnarray}
as a function of $k$ is shown in the middle panel of Fig. \ref{figure_cor_EHM}. The maximum of the correlation function occurs at $k=\pi$. We can similarly deduce that the dominating configuration are the spin-density wave (SDW) states $|\uparrow,\downarrow,\cdots,\uparrow,\downarrow\rangle$ and $|\downarrow,\uparrow,\cdots,\downarrow,\uparrow\rangle$. These results obtained are consistent with previous studies \cite{EHM}.

\subsection{Case III: $V\ne 0$, $U=4$}\label{sec:EHM2}
Let's now consider a block size of two sites, i.e. $N_B=2$. In the
following, we will use $i$ to denote a single-site and $\tilde{i}$
to denote a single-block consist of two neighboring sites. The single-block
(two-site) reduced density matrix $\rho_{\tilde{i}}$ is calculated
for the case of $U=4$ and $N=14$. In the basis of
$\{\ket{0},\ket{\uparrow},\ket{\downarrow},\ket{\uparrow\downarrow}\}\otimes\{\ket{0},\ket{\uparrow},\ket{\downarrow},\ket{\uparrow\downarrow}\}$,
the eigenstates of $\rho_{\tilde{i}}$ have the form of
\begin{eqnarray}
|\phi_A\rangle&=&\vert\uparrow\downarrow,\uparrow\downarrow\rangle,\nonumber\\
|\phi_B\rangle&=&|0,0\rangle,\nonumber\\
|\phi_C\rangle&=&\frac{1}{\sqrt{2}}\big(\vert\uparrow\downarrow,\downarrow\rangle+\vert\downarrow,\uparrow\downarrow\rangle\big),\nonumber\\
|\phi_D\rangle&=&\frac{1}{\sqrt{2}}\big(\vert\uparrow,0\rangle-\vert0,\uparrow\rangle\big),\nonumber\\
|\phi_E\rangle&=&\frac{1}{\sqrt{2}}\big(\vert\downarrow,0\rangle-\vert0,\downarrow\rangle\big),\nonumber\\
|\phi_F\rangle&=&\frac{1}{\sqrt{2}}\big(\vert\uparrow\downarrow,\uparrow\rangle+\vert\uparrow,\uparrow\downarrow\rangle\big),\nonumber\\
|\phi_G\rangle&=&\alpha\big(\vert\uparrow\downarrow,0\rangle+\vert0,\uparrow\downarrow\rangle\big)-\beta\big(\vert\uparrow,\downarrow\rangle-\vert\downarrow,\uparrow\rangle\big),\nonumber\\
|\phi_H\rangle&=&\frac{1}{\sqrt{2}}\big(\vert\uparrow,\downarrow\rangle+\vert\downarrow,\uparrow\rangle\big),\nonumber\\
|\phi_I\rangle&=&\vert\uparrow,\uparrow\rangle,\nonumber\\
|\phi_J\rangle&=&\vert\downarrow,\downarrow\rangle,\nonumber\\
|\phi_K\rangle&=&\frac{1}{\sqrt{2}}\big(\vert\uparrow\downarrow,\downarrow\rangle-\vert\downarrow,\uparrow\downarrow\rangle\big),\nonumber\\
|\phi_L\rangle&=&\frac{1}{\sqrt{2}}\big(\vert\uparrow,0\rangle+\vert0,\uparrow\rangle\big),\nonumber\\
|\phi_M\rangle&=&\frac{1}{\sqrt{2}}\big(\vert\downarrow,0\rangle+\vert0,\downarrow\rangle\big),\nonumber\\
|\phi_N\rangle&=&\frac{1}{\sqrt{2}}\big(\vert\uparrow\downarrow,\uparrow\rangle-\vert\uparrow,\uparrow\downarrow\rangle\big),\nonumber\\
|\phi_O\rangle&=&\frac{1}{\sqrt{2}}\big(\vert\uparrow\downarrow,0\rangle-\vert0,\uparrow\downarrow\rangle\big),\nonumber\\
|\phi_P\rangle&=&\gamma\big(\vert\uparrow\downarrow,0\rangle+\vert0,\uparrow\downarrow\rangle\big)+\delta\big(\vert\uparrow,\downarrow\rangle-\vert\downarrow,\uparrow\rangle\big),
\label{eq:EVec_EHM}
\end{eqnarray}
where $\alpha,\beta,\gamma,\delta$ are positive real numbers.

\begin{figure}[tbp]
\centering
\includegraphics[width=8.5cm]{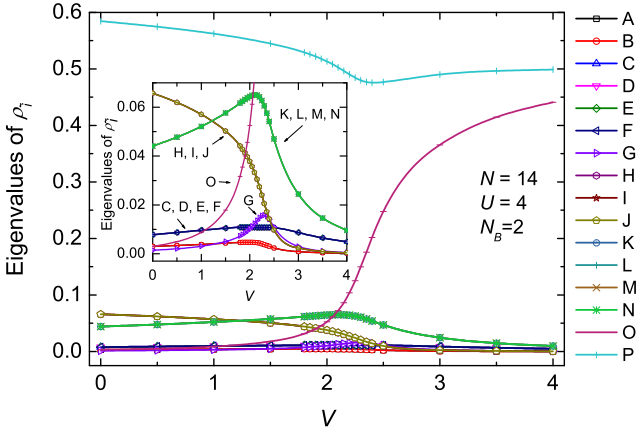}
\caption{(color online) The eigenvalues of the single-block (two-site) reduced
density matrix $\rho_{\tilde{i}}$ as a function of $V$ in the 1D
extended Hubbard model. $N=14$ and $U=4$. The inset shows a close-up
of the low-lying states.} \label{fig:RDM2_EHM}
\end{figure}

\begin{figure}[tbp]
\centering
\includegraphics[width=8cm]{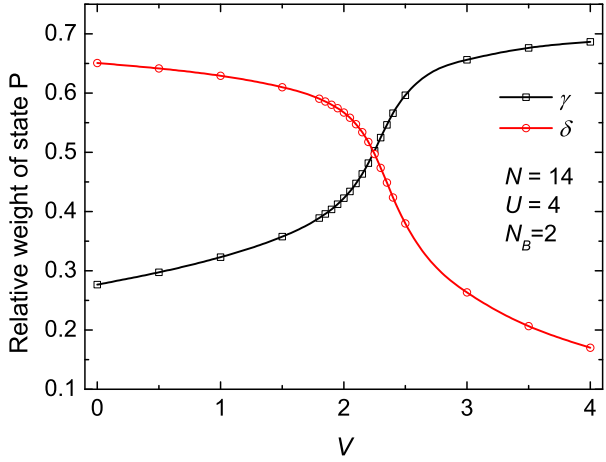}
\caption{(color online) The relative weight of the $\gamma$ and $\delta$ in the eigenstate $\ket{\phi_P}$ of the single-block reduced density matrix as a function of $V$. $N=14$, $U=4$, and $N_B=2$.} \label{fig:RDM_P_EHM}
\end{figure}

Figure \ref{fig:RDM2_EHM} shows a plot of the eigenvalues of the
corresponding eigenstates as a function of $V$. From the figure, we
can see that some of the states, for examples, states
$\{\ket{\phi_C},\ket{\phi_D},\ket{\phi_E},\ket{\phi_F}\}$ and states
$\{\ket{\phi_K},\ket{\phi_L},\ket{\phi_M},\ket{\phi_N}\}$, are
degenerated respectively. The degeneracy tells us that there are
spin-up-down and charge symmetries in the system.

Among all the eigenstates, the weight of the state $\ket{\phi_P}$ is
dominated for the whole range of value of $V$ shown. This state is a
spin-singlet state. It is a superposition of two parts, the charge
part $\ket{\uparrow\downarrow,0}+\ket{0,\uparrow\downarrow}$ and the
spin part $\ket{\uparrow,\downarrow}+\ket{\downarrow,\uparrow}$ with
relative weight characterized by $\gamma$ and $\delta$ respectively.
The magnitude of $\gamma$ and $\delta$ are plotted as a function of
$V$ in Fig. \ref{fig:RDM_P_EHM}. There is a crossing between the two
magnitudes around $V=2$. For $V>>2$, the relative weight of the
charge part is much greater than that of the spin part. Besides,
considering the region $V>>2$ in Fig. \ref{fig:RDM2_EHM} again, the
second dominating state is $\ket{\phi_O}$ which overwhelms all other
eigenstates except $\ket{\phi_P}$. From Eq. (\ref{eq:EVec_EHM}), we
also notice that the state $\ket{\phi_O}$ only consists of the
charge part. As a result, we may argue that in the this region, the
charge part is decoupled from the spin part. The reduced density
matrix can be reduced to an effective one in Eq. (\ref{eq: rho_charge}).

Similarly, for $V<<2$, the spin part in $\ket{\phi_P}$ outweigh the
charge part as one can realize from Fig. \ref{fig:RDM_P_EHM}. The
eigenvalues for $\ket{\phi_H}$, $\ket{\phi_I}$, and $\ket{\phi_J}$
also dominate in the low-lying eigenstates. The spin degree of
freedom is singled out in this regime and the reduced density matrix
can be effectively expressed as the one in Eq. (\ref{eq:singlerdm}).

\begin{figure}[tbp]
\centering
\includegraphics[width=8cm]{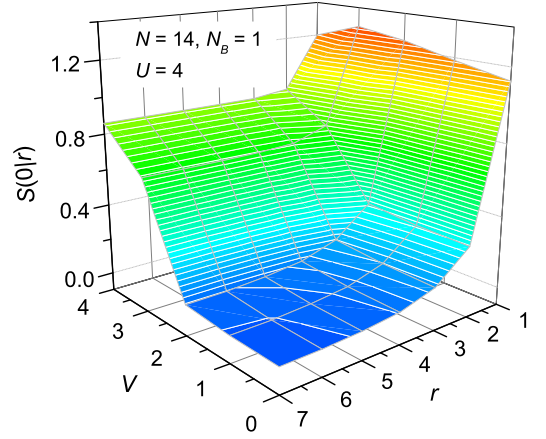}\\\vspace{20pt}
\includegraphics[width=8cm]{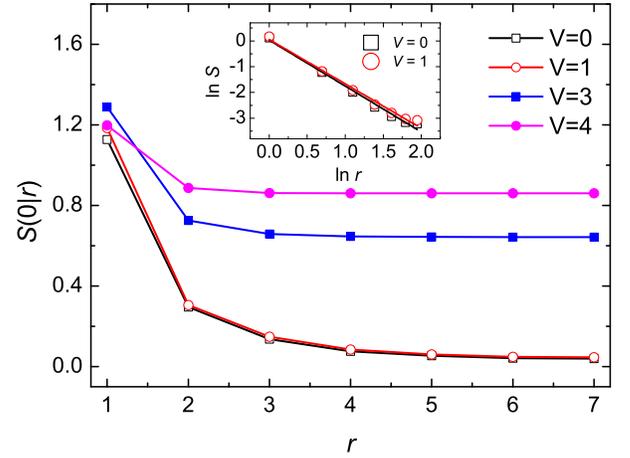}
\caption{(color online) (Top) A 3D plot of the mutual information between two
single sites as a function of $V$ and $r$ in the 1D extended Hubbard
model. $N=14$, $U=4$ and $N_B=1$. (Bottom) A plot of the mutual
information as a function of $r$ for $V=0,1,3,4$. The inset shows a
ln-ln plot of the mutual information as a function of $r$ for
$V=0,1$.} \label{fig:3D_entro_EHM_U4}
\end{figure}


\begin{figure}[tbp]
\centering
\includegraphics[width=8cm]{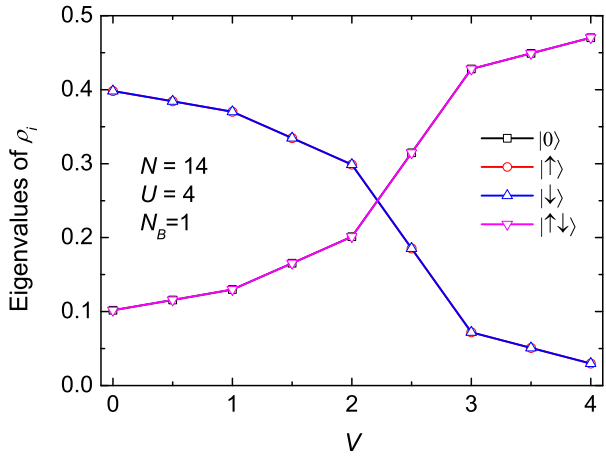}
\caption{(color online) The eigenvalues of the single-site reduced density matrix
as a function of $V$ in the 1D extended Hubbard model. Here $N=14$,
$U=4$ and $N_B=1$.} \label{fig:EVRDM_EHM_U4}
\end{figure}

In Fig. \ref{fig:3D_entro_EHM_U4}, the mutual information between
two single sites are calculated as a function of $V$ and the
separation $r$. In the 2D plot, we can see that the mutual
information for $V>2$ is non-vanishing at a long distance. For
$V<2$, the ln-ln plot in the inset also shows that the mutual
information decays algebraically with $r$. So the long-range
correlation in the two regimes is already captured by a block size
of one site. To construct the order parameter in these regimes, we
may go back to consider the single-site reduced density matrix.

Figure \ref{fig:EVRDM_EHM_U4} is a plot of the eigenvalues of the
single-site reduced density matrix. It is qualitatively the same as
the case of $U=6$. The $\ket{\uparrow}$ and $\ket{\downarrow}$
states are dominated for $V<2$ while the $\ket{0}$ and
$\ket{\uparrow\downarrow}$ states are dominated for $V>2$. Following
similar analysis in the previous sections, we can find that the
order operators are $\sigma_i^z$ and $\eta_i^z$, which characterize
the SDW and CDW respectively, in the two regions.

However, the above is not the whole story. The single-site reduced
density matrix was not enough to capture the correlation in the
system around $V=2$. Returning to Fig. \ref{fig:RDM2_EHM}, the
eigenvalue of $\ket{\phi_P}$ has a drop around $V=2$ and there is
also a relatively large rise in the weight of the eigenstates
$\ket{\phi_K}$, $\ket{\phi_L}$, $\ket{\phi_M}$ and $\ket{\phi_N}$.
In addition, the magnitude of $\gamma$ and $\delta$ in
$\ket{\phi_P}$ becomes comparable in this intermediate value. These
observation suggest that the spin part and the charge part are
coupled around $V=2$. There may exist some other kind of long-range
correlation rather than SDW and CDW in this intermediate region.

Let's consider $N_B=2$ again and rearrange the single-block reduced
density matrix from the basis of
$\{\ket{0},\ket{\uparrow},\ket{\downarrow},\ket{\uparrow\downarrow}\}_i\otimes\{\ket{0},\ket{\uparrow},\ket{\downarrow},\ket{\uparrow\downarrow}\}_{i+1}$
to
$\{\ket{0,0},\ket{0,\uparrow},\ket{\uparrow,0},\ket{\uparrow,\uparrow}\}_{\tilde
i}\otimes\{\ket{0,0},\ket{0,\downarrow},\ket{\downarrow,0},\ket{\downarrow,\downarrow}\}_{\tilde
i}$. The purpose of doing this is that we want to filter out the
spin-down degree of freedom from the spin-up degree of freedom (or
vice versa). After the rearrangement, we traced out the spin-down
degree of freedom. The resulting reduced density matrix has the form
of
\begin{eqnarray}
\rho_{\tilde i\uparrow}=\begin{pmatrix}
    u & 0 & 0 & 0 \\
    0 & v & z & 0 \\
    0 & z & v & 0 \\
    0 & 0 & 0 & u \\
  \end{pmatrix},
\end{eqnarray}
in the basis of
$\{\ket{0,0},\ket{0,\uparrow},\ket{\uparrow,0},\ket{\uparrow,\uparrow}\}_{\tilde
i}$. Similarly, if we trace out the degree of freedom for spin-up,
we have
\begin{eqnarray}
\rho_{\tilde i\downarrow}=\begin{pmatrix}
    u & 0 & 0 & 0 \\
    0 & v & z & 0 \\
    0 & z & v & 0 \\
    0 & 0 & 0 & u \\
  \end{pmatrix},
\end{eqnarray}
in the basis of
$\{\ket{0,0},\ket{0,\downarrow},\ket{\downarrow,0},\ket{\downarrow,\downarrow}\}_{\tilde
i}$.

Next, we would like to compare the off-diagonal matrix elements $z$
in $\rho_{\tilde i\uparrow}$ and $\rho_{\tilde i\downarrow}$ with
that corresponding to SDW and CDW. The weight of SDW is given by the
coefficient of $\ket{\uparrow,\downarrow}\bra{\uparrow,\downarrow}$
or $\ket{\downarrow,\uparrow}\bra{\downarrow,\uparrow}$ (the
notation here is in form of $\oper{i,i+1}{i,i+1}$). From Eq.
(\ref{eq:EVec_EHM}), it can be obtained by
\begin{eqnarray}
P_{\text{SDW}}=\delta^2p_P+0.5p_H+\beta^2p_G,
\end{eqnarray}
where the $p$'s are the eigenvalues of the corresponding eigenstates
in Eq. (\ref{eq:EVec_EHM}). For the weight of CDW, we obtained it by
considering the matrix element corresponds to
$\ket{0,\uparrow\downarrow}\bra{0,\uparrow\downarrow}$ or
$\ket{\uparrow\downarrow,0}\bra{\uparrow\downarrow,0}$ from Eq.
(\ref{eq:EVec_EHM}). We have
\begin{eqnarray}
P_{\text{CDW}}=\gamma^2p_P+0.5p_O+\alpha^2p_G.
\end{eqnarray}

\begin{figure}[tbp]
\centering
\includegraphics[width=8cm]{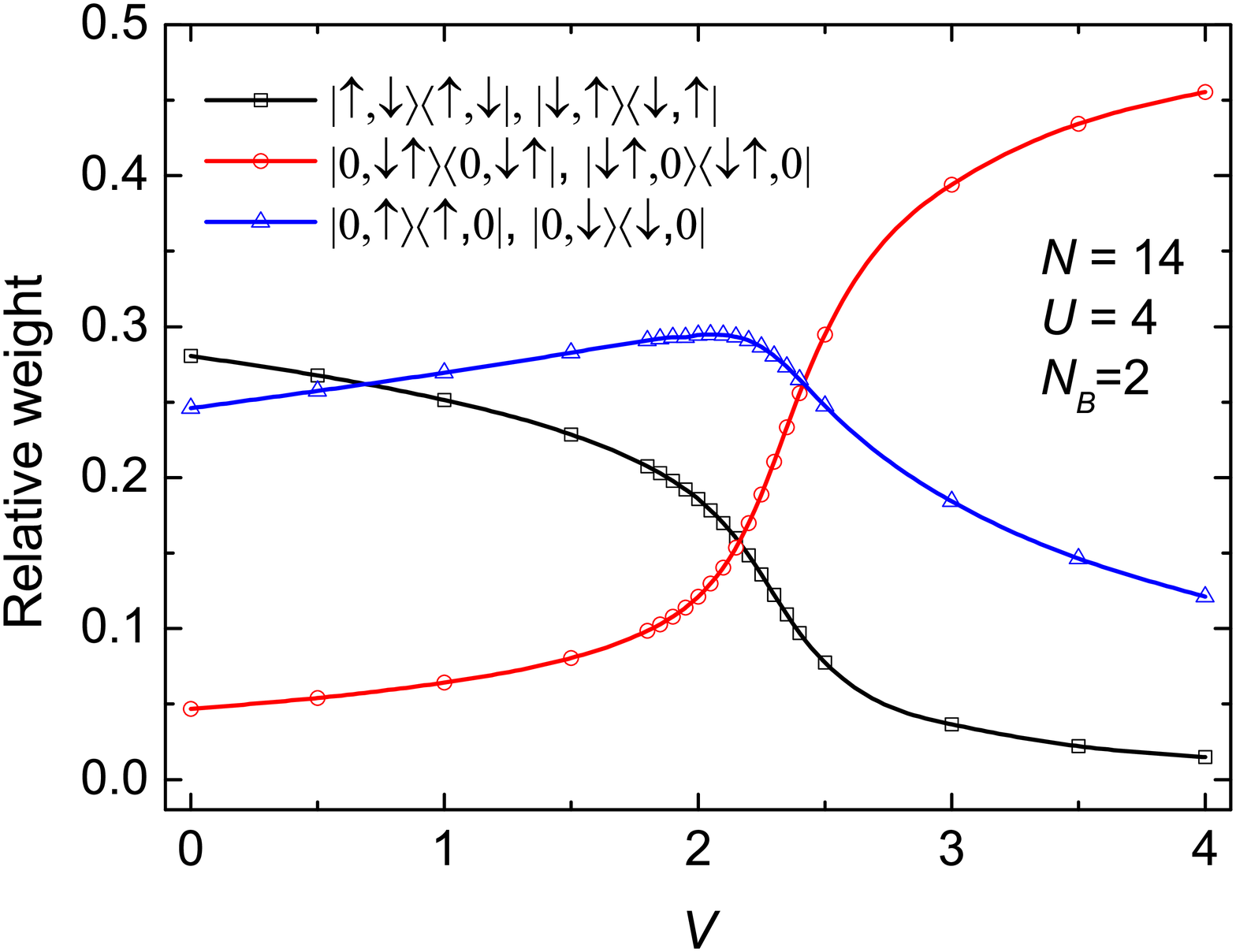}
\caption{(color online) The relative weight of some of the matrix elements in the
single-block reduced density matrix in the 1D extended Hubbard model
as a function of $V$. $N=14$, $U=4$ and $N_B=2$.}
\label{fig:RDM_order_EHM}
\end{figure}

Figure \ref{fig:RDM_order_EHM} shows a plot of $P_{\text{SDW}}$,
$P_{\text{CDW}}$ and $z$ as a function of $V$. On the two sides far
away from $V=2$, the largest weight would correspond to SDW and CDW
respectively as expected from previous analysis. However, around
$V=2$, the off-diagonal matrix element in $\rho_{\tilde i\uparrow}$
and $\rho_{\tilde i\downarrow}$ are dominating instead. If we just
pick this dominating weight to define the order parameter, we have
\begin{eqnarray}
O_{\tilde
i}&=&\omega_1\oper{0,\uparrow}{\uparrow,0}+\omega^*_1\oper{\uparrow,0}{0,\uparrow}\nonumber\\
&&+\omega_2\oper{0,\downarrow}{\downarrow,0}+\omega^*_2\oper{\downarrow,0}{0,\downarrow}.
\end{eqnarray}
As mentioned before, the system possess up-down spin symmetry. We
could take $\omega_1=\omega_2=\omega$, and then separate the
operator into the real and imaginary part. We have
\begin{eqnarray}
O_{\tilde
i}&=&\omega\Big(\oper{0,\uparrow}{\uparrow,0}+\oper{0,\downarrow}{\downarrow,0}\Big)\nonumber\\
&&+\omega^*\Big(\oper{\uparrow,0}{0,\uparrow}+\oper{\downarrow,0}{0,\downarrow}\Big),\nonumber\\
&=&x\Big(\oper{0,\uparrow}{\uparrow,0}+\oper{0,\downarrow}{\downarrow,0}\nonumber\\
&&+\oper{\uparrow,0}{0,\uparrow}+\oper{\downarrow,0}{0,\downarrow}\Big)\nonumber\\
&&+iy\Big(\oper{0,\uparrow}{\uparrow,0}+\oper{0,\downarrow}{\downarrow,0}\nonumber\\
&&-\oper{\uparrow,0}{0,\uparrow}-\oper{\downarrow,0}{0,\downarrow}\Big).
\end{eqnarray}
Either the first term or the second term in the bracket above, or
their linear combination can be taken as the order parameter. Let's
take the real part as the order parameter, i.e
\begin{eqnarray}
O_{\tilde
i}&=&\oper{0,\uparrow}{\uparrow,0}+\oper{0,\downarrow}{\downarrow,0}\nonumber\\
&&+\oper{\uparrow,0}{0,\uparrow}+\oper{\downarrow,0}{0,\downarrow}.
\end{eqnarray}
In terms of the fermion creation and annihilation operators, we have
\begin{eqnarray}
O_{\tilde
i}=c_{i,\uparrow}^{\dagger}c_{i+1,\uparrow}+c_{i+1,\uparrow}^{\dagger}c_{i,\uparrow}+c_{i,\downarrow}^{\dagger}c_{i+1,\downarrow}+c_{i+1,\downarrow}^{\dagger}c_{i,\downarrow},
\end{eqnarray}
which is the conventional order parameter that has been used to
investigate the BOW in the extended Hubbard model \cite{Zhang2004}.

\section{summary}
\label{sec:sum}
We summarize our scheme of constructing the order parameter in the following:
(i) Calculate the mutual information to see whether there exists any long-range correlation in the system. Choose the minimum size of the block where the mutual information is non-vanishing for two blocks separated by a large distance.
(ii) Calculate the difference between the diagonal matrix elements of the two block reduced density matrix and the product of the diagonal matrix elements of two single block reduced density matrix. This is to single out the correlating elements from other noise due to the finite size effect.
(iii) Obtain the eigenvalues and eigenvectors of the single block reduced density matrix. Construct the order parameter from the heavy weighted eigenstates. Apply traceless condition and cut-off condition to narrow down the choice.
(iv) Calculate the correlation function using the derived order parameter and study the mode of it.

With the scheme proposed, we have derived the order parameters which show
long-range correlation in the ground state of the 1D EHM
without using any empirical knowledge. Such an application confirmed
that the order parameter for a quantum many-body system can be
systematically derived even without the knowledge of symmetry in the
system. We expect that our scheme can shed new lights on the
controversies in some frustrated antiferromagnet\cite{frustraedSpin}
and bond-order wave in the extended Hubbard
model\cite{BOW, Jeckelmann,Jeckelmann2}.

Throughout the paper, the system size being simulated was 14
sites. Remarkably, we see that even for such a small system size,
the spectra of the single-site or two-site reduced density matrices
were able to capture the information of the nature of the order in
the ground state of the system. This may be a promising scheme as
one could have insight on the order existing in the system without
simulating a large system, which requires intensive computational
powers. However, we would like to mention that the crossing of the
dominating states in the reduced density matrix spectrum may not
locate the phase boundaries exactly. To locate the phase boundaries,
one can use the order operator derived to calculate the correlation function and then perform finite size scaling analysis.

This work is supported by the Earmarked Grant Research from the
Research Grants Council of HKSAR, China (Project No. CUHK 401213).

\end{document}